\documentstyle[psfig]{mn}
\newif\ifAMStwofonts

\newcommand{\be}{\begin{equation}}
\newcommand{\ee}{\end{equation}}
\newcommand{\bea}{\begin{eqnarray}}
\newcommand{\eea}{\end{eqnarray}}
\def\lesssim{\,\lower2truept\hbox{${<\atop\hbox{\raise4truept\hbox{$\sim$}}}$}\,}
\def\gtrsim{\,\lower2truept\hbox{${>\atop\hbox{\raise4truept\hbox{$\sim$}}}$}\,}

\title[Joint Formation of QSOs and Spheroids]{Joint Formation of QSOs and Spheroids:
QSOs as clocks of star formation in Spheroids}

\author[G.L.\ Granato et al.]
  {G.L.\ Granato,$^{1,2}$ L.\ Silva$^{3,2}$, P.\ Monaco$^{4}$, P.\ Panuzzo$^2$,
  P.\ Salucci$^2$, G.\ De Zotti$^{1,2}$, 
\newauthor
  L.\ Danese$^2$\\
$^1$Osservatorio Astronomico di Padova, Vicolo Osservatorio 5, I35122 Padova, Italy\\
$^2$SISSA, via Beirut 2-4, I34014 Trieste, Italy\\
$^3$Osservatorio Astronomico di Trieste, via Tiepolo 11, I34131 Trieste, Italy\\
$^4$Dipartimento di Astronomia, via Tiepolo 11, I34131 Trieste, Italy}

\date{Accepted....
      Received....
      in original form....}
\pagerange{\pageref{firstpage}--\pageref{lastpage}}
\pubyear{}
\begin{document}
\label{firstpage}
\maketitle
\begin{abstract}
Direct and indirect observational evidence leads to the conclusion
that high redshift QSOs did shine in the core of early type
proto--galaxies during their main episode of star formation. Exploting
this fact, we derive the rate of formation of this kind of stellar
systems at high redshift by using the QSO Luminosity Function. The
elemental proportions in elliptical galaxies, the descendents of the
QSO hosts, suggest that the star formation was more rapid in more
massive objects. We show that this is expected to occur in Dark Matter
haloes, when the processes of cooling and heating is considered. This
is also confirmed by comparing the observed sub--mm counts to those
derived by coupling the formation rate and the star formation rate of
the spheroidal galaxies with a detailed model for their SED evolution.
In this scenario SCUBA galaxies and Lyman Break Galaxies are early
type proto--galaxies forming the bulk of their stars
before the onset of QSO activity.  
\end{abstract}


\section{Introduction}
\label{sec:int}

QSOs had  been for a long time the main probe of the epoch when galaxies are
thought to have formed. Hubble Space Telescope observations, especially the
HDF surveys, and ground based observations with the new 10 meter class
telescopes, have opened to direct investigation also the early phases of
galaxy formation. Exploration of the local Universe is also yielding
extremely relevant results to understand the QSO phenomenon. Very high
angular resolution photometric and spectroscopic observations have
demonstrated that Massive Dark Objects (MDO) are generally present in nearby
galaxies endowed with significant spheroidal components (Magorrian et al.\
1998; van der Marel 1999). MDOs are thought to be dormant BHs which spent
their shining phase as QSOs. Indeed, the MDO mass function well matches the
mass function of baryons accreted onto BHs during the QSO activity (Salucci
et al.\ 1999, hereafter paper I). Estimated MDO masses are roughly
proportional to those of the spheroidal component of the host galaxy
(Magorrian et al.\ 1998; Gebhardt et al.\ 2000) . This fact implies that
high redshift and highly luminous QSOs have been hosted in massive early
type galaxies.

In addition, spectroscopic observations of QSO emission  and absorption
lines show that high redshift QSOs live in a metal enriched environment (see
Hamann \& Ferland 1999 for a comprehensive review). Observations at far-IR
and sub-mm wavelengths uncovered large amounts of dust in QSOs (Omont et
al.\ 1996; Ivison et al.\ 1998; Benford et al.\ 1999; Yun et al.\ 2000). The
dust emission has been ascribed to starburst in the host galaxy (see e.g.\
Yun et al.\ 2000; Chapman et al.\ 1998; Benford et al.\ 1999) or dust
illuminated by the active nucleus (see e.g.\ Andreani, Franceschini \&
Granato 1999). The evidence of the dust illuminated by starbursts in high
redshift radio galaxies is solid (see e.g. Archibald et al.\ 2000).

All these pieces of evidence suggest that QSOs at high redshift are active
nuclei shining in early type galaxies, during the short fraction of the
Hubble time when they were vigorously forming stars and still gas rich. The
high metal abundance of the QSO environments strongly supports the idea that
the bulk of star formation in host galaxies occurred before the QSO shining
phase.

In fact there is evidence that the bulk of stars have formed very soon in
elliptical galaxies and in the prominent bulges of the spirals. Cluster elliptical
galaxies exhibit a very tight color-magnitude relation, which can be explained if
the bulk of their stars formed at very early epochs, corresponding to $z\gtrsim 2$
(Bower et al.\ 1992; Ellis et al.\ 1997; Kodama et al.\ 1998). The very weak
dependence of the $Mg_2-\sigma$ relation on the galaxy environment suggests that
field ellipticals are on average at most 1 Gyr younger than cluster ellipticals
(Bernardi et al.\ 1998). The bulges of spiral galaxies exhibit a tight correlation
between the $Mg_2$ index and the intrinsic luminosity similar to that of the
ellipticals (Jablonka, Martin \& Arimoto 1996). These facts support the conclusion
that spheroids formed stars very rapidly at early epochs (see e.g.\ Renzini 1999).
The rapid star formation can be strongly reduced or halted by the exhaustion of
the gas in most massive objects or by heating and winds, if SN explosions (or the
QSO activity itself, as argued below) transfer to ISM even a fraction of their
energy (see e.g.  Dekel and Silk 1986). For instance galactic winds were brought
in the play in order to explain the mass-metallicity relation in elliptical
galaxies ( Mathews \& Baker 1971; Larson 1974). The observed iron abundances in
groups and clusters of galaxies can be explained if galactic winds expelled large
amounts of enriched gas from spheroids (see e.g.\ Renzini 1997). Many photometric
studies confirm that the massive early type galaxies were practically formed at
$z\simeq 1$ with a number density very close to the local one (Im et al.\ 1996;
Franceschini et al.\ 1998; Schade et al.\ 1999), even though the issue is still
somewhat controversial (e.g.\ Kauffmann, Charlot \& White, 1996; Fontana et al.
1999).

In this paper we present evidence in favour of a strict relationship in the
formation of QSOs and their spheroid hosts, with particular emphasis on the
timing of star formation in the hosts and the QSO shining. We derive the
formation rate of the spheroids from the formation rate of the QSOs (sect
2.1 and 2.2), taking into account the observational and theoretical supports
for higher star formation rates (SFRs) in more massive spheroids.  In
section 3 we present the model (GRASIL) we use to estimate the evolution of
the spectral energy distribution (SED) of the spheroids. The model includes
dust. In section 4, exploting the formation rate of spheroids coupled with
GRASIL, we argue that the QSO hosts at $z\geq 2$ are the galaxies uncovered
in the submillimetre surveys with SCUBA (Blain et al.\ 1999a,b,c). In
section 5 we discuss relationship of the young spheroids with Lyman Break
Galaxies (Steidel, Pettini and Hamilton 1995; Steidel et al.\ 1996; Madau et
al.\ 1996) and in section 6 we investigate the QSO phase.  In section 7 we
place our results in the context of hierarchical cosmogonies. Section 8 is
devoted to a summary.

Unless otherwise specified, the results we present have been obtained
adopting a cosmological model with $H_0 = 70\ \mbox{km}\ \mbox{s}^{-1}\
\mbox{Mpc}^{-1}$, $\Omega_\Lambda=0.7$, and $\Omega_M=0.3$. We also
performed the same computations in an Einstein-de Sitter cosmology with $H_0
= 50$. The results, being essentially identical, are not  shown. To avoid
confusion, we indicate with lowercase $t$, times measured from the Big Bang,
while galactic ages (i.e.\ times measured from the onset of star formation
in a galaxy) are indicated with uppercase $T$.

\section{The formation of spheroidal galaxies as traced by QSO evolution}
\label{sec:papII}

The epoch-dependent QSO Luminosity Function (LF) $n_{\rm QSO}(L_{\rm
QSO},z)$ is well defined in a large redshift interval $0\leq z\leq 3$ both
in the optical and in the soft X-ray bands (Boyle et al.\ 1993; Pei 1995;
Kennefick et al.\ 1996; Grazian et al.\ 2000; Miyaji, Hasinger and Schmidt
2000); also new optical  surveys allow sound estimates of the QSO evolution
at higher redshift. (Fan et al.\ 2000). Observations with Chandra X--ray
Observatory have shown that most of the Hard X-ray Background (HXRB) is due
to active nuclei ( Mushotzky et al 1999; Giacconi et al 2000; Barger et al
2000). In the following section we will use the Luminosity Function of QSOs,
including the 'obscured ' ones responsible for a large fraction of the HXRB,
in order to infer the formation rate of the host spheroidal galaxies.

\subsection {The formation rate of spheroids}

Since quasar lifetimes $\Delta t_{\rm Q}$ are short compared to the
typical evolution timescale of their LFs, the rate at which QSOs with BH mass
$M_{\bullet}$ shine at time $t_{QSO}$ (corresponding to redshift $z_{QSO}$) is
related to the LF by:
\be \dot n_\bullet (M_\bullet (L_{\rm QSO}),t_{QSO}) =
\frac{n_{\rm QSO}(L_{\rm QSO},z_{QSO})}{\Delta t_{\rm Q}}\frac{dL_{\rm QSO}}
{dM_{\bullet}} 
\label{eq:life}
\ee

\noindent 
Here $n_{\rm QSO}$ is taken from Pei (1995) for the optical bands,
while the contribution from obscured objects is estimated using the LF
of Boyle et al. (1993) (see Comastri et al. 1995 and paper I for
details).  The luminosity of a QSO in a given e.m. band is related to
the central BH mass $M_\bullet$ through: \be L_{\rm QSO}(M_\bullet)
=f_{\rm ED} L_{\rm ED}(M_\bullet) /C_B, \label{eq:lqso} \ee

\noindent where $C_B$ is the bolometric correction appropriate for the e.m.
band used and  $f_{\rm ED} \equiv L_{\rm bol}/L_{\rm ED}$ is the bolometric
luminosity in units of the Eddington luminosity. The available data on AGN and
QSOs suggest that the ratio $f_{\rm ED}$ is a function of the redshift and/or
of the luminosity, going from $\sim 0.05 - 0.1$ for faint local AGNs, to
$\gtrsim 1$ for bright high redshift QSOs (see, e.g., Padovani 1989; 
Sun and Malkan 1989; Wandel 1999). This behaviour can be represented 
as function of the luminosity by:

\be f_{\rm ED} = \left(\frac{L_{\rm bol}}{10^{49} {\rm
erg/s}}\right)^{\alpha_{\rm ED}},
\label{eq:edeff} \ee

\noindent
with the exponent $\alpha_{\rm ED}$ set to 0.2 (cfr paper I).

\noindent We further assume that quasar lifetimes are proportional
to $t_{\rm duty}$, the e--folding time for the exponential
growth of BH mass during QSO activity:
\be  \Delta t_{\rm Q} = K t_{\rm duty} = K \varepsilon\, {M_{\bullet} c^2\over L_{\rm
ED}f_{\rm ED}}. \label{eq:duty} \ee

\noindent If we adopt $\epsilon=0.1$ for the mass to radiation conversion
efficiency of accretion, we have  $t_{\rm duty}=4 \times 10^7 $ yr for 
$f_{ED}=1$.
The value $\Delta t_Q $  can be inferred by matching the local BH Mass Function
with the Mass Function inferred from accretion (Salucci et al 1999),
by the QSO clustering properties (Martini and Weinberg 2000)
or by matching the space density of high redshift QSOs and the
space density of local massive spheroids (Richstone et al 1998; Monaco et 
al 1999). For instance the analyses of Martini and Weinberg (2000),
Salucci et al (1999) and Monaco et al (2000) suggest
$ \Delta t_{\rm Q} \sim 8 \times 10^6-10^8 $ yr.

\noindent Using equations (\ref{eq:lqso}) and (\ref{eq:edeff}) into equation
(\ref{eq:life}) we get:

\bea
\lefteqn{\dot n_\bullet (M_\bullet,t_{QSO})=} \nonumber  \\ &&
\label{eq:qso}
n_{\rm QSO}(L_{\rm QSO}(M_\bullet),z_{QSO})
\frac{L_{\rm QSO}(M_\bullet)}{M_\bullet}\frac{1}{(1-\alpha_{\rm ED})}\,
\frac{1}{ \Delta t_{\rm Q}} \; .   \eea

\noindent 
In order to proceed we introduce here a major assumption, namely that
the relationship between the BH mass and the host mass (or velocity
dispersion), holding for local spheroids (Magorrian et al 1998;
Ferrarese and Merritt 2000; Gebhardt et al 2000), has been imprinted
during the early phase of the QSO and host evolution (see e.g. Silk
and Rees 1998; Fabian 1999), and has not been changed significantly by
subsequent mergers. Then $\dot n_\bullet (M_\bullet,t_{QSO})$ is
linked to the rate $\dot n_{sph}(M_{sph},t_{QSO})$ at which galaxies
endowed with spheroid of present day mass $M_{sph}$ appear to host a
QSO at time $t_{QSO}$ by, \be \dot n_\bullet (M_\bullet, t_{QSO})=
\int \dot n_{sph}(M_{sph},t_{QSO}) f(x) dx ,
\label{eq:esph} \ee

\begin{figure}
\psfig{file=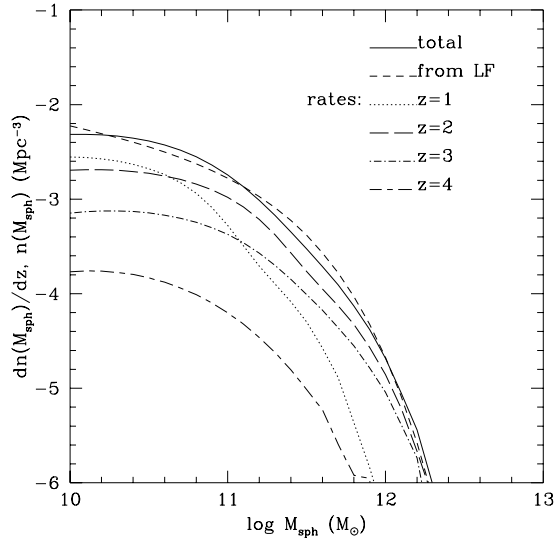,width=7.5truecm}
\caption{Shining rate of bulges $dn(M_{sph}/dz$ at redshift 1 (dotted thin
line), 2 (long-dashed), 3 (dot-dashed) and 4 (short-long dashed).  The
thick continuous line shows the total mass function of bulges, $n(M_{sph})$,
compared to the one inferred in paper I from the local LF of galaxies
(thick dashed line).
}
\label{fig:bulges}
\end{figure}

\noindent where $f(x)$ is the distribution of
$x=log(\frac{M_{\bullet}}{M_{sph}})$. Data on local objects suggest that
$f(x)$ can be represented by a gaussian with mean $\langle x \rangle =-2.6$
and $\sigma=0.3$ (see paper I).  Equating the second members of the two
equations above, we get a relationship which, once deconvolved, yield an
estimate of $\dot n_{sph}(M_{sph},t_{QSO})$, except for the somewhat
uncertain factor $1/(K \varepsilon)$. This residual
uncertainty can be eliminated
normalizing the result in order to recover the local LF of spheroids.
Fig.~\ref{fig:bulges} shows that there is a good agreement between predictions
and observations, provided that $K \varepsilon\ \simeq 0.1$.
In the figure the rate
$\dot n_{sph}(M_{sph},t_{QSO})$ at different redshift is presented.
Integration of $\dot n_{sph}(M_{sph},t_{QSO})$ over time yields their
present day mass function. To compare it with data, we use the following
relationship between the visible mass of a spheroid $M_{\rm sph}$ and its
light: 
\be M_{\rm sph}/L=(M/L)_{*}(L/L_{*E})^ {\beta_{\rm sph}-1},
\label{eq:msuelle} \ee
\noindent where $L_{*E}$ is the Schechter parameter for the E luminosity
function, $\beta_{\rm sph}=1.25$ and $(M/L)_{*}= 4.9$ for luminosities  in the
$B$ band (cfr. paper I). With  this recipe the local LF of  spheroids is
translated into a Mass Function (see e.g. Salucci et al.\ 1999).

We are mainly interested to high redshift, since we want to test the
hypothesis that the QSO activity signals the end of the main episode of star
formation in spheroids; in our view the low redshift $z\leq 1$ activity is
mostly due to reactivation (see e.g. Kauffman and Haehnelt 2000; Cavaliere
and Vittorini 2000). The galaxies associated with QSOs are bright objects.
Following the correlation between the bulge luminosity and the BH mass
(Magorrian et al 1998) or the correlation between velocity dispersion and BH
mass (Ferrarese and Merritt 2000; Gebhardt et al 2000 ), QSOs, which
typically have $L_{bol}\gtrsim 3\times 10^{45}$ and  can be associated to
massive BH ($M_{BH}\gtrsim 2\times 10^7$), are expected to have been hosted in
spheroidal galaxies with central velocity dispersion $\sigma \gtrsim 100$
km/s and with present luminosity $M_B \lesssim -18.$

\subsection {Early star formation in spheroidal galaxies}

In the Introduction we have reviewed several pieces of evidence
suggesting that star formation begins in the hosting spheroids at a time
$t_{\star}$ and proceeds vigoursly at least until the time  $t_{QSO}$, when
the QSO shines. Here we estimate the duration of the star formation phase
and its possible dependence on the mass.

The statistics on Broad Emission Lines suggest that metallicities are possibly
higher in more luminous QSOs (Hamann and Ferland 1993, 1999). This is
reminiscent of the mass-metallicity relationship in the local ellipticals and
can be naturally explained taking into account the correlation between QSOs and
host galaxy luminosities expected on the basis the observed correlation between
BH remnants and spheroidal masses.

Large elliptical galaxies exhibit, at least in the central regions,
overabundances of magnesium and possibly of other $\alpha$-elements. Fisher
et al (1995) have demonstrated that ellipticals with larger central velocity
dispersion $\sigma_0$ tend to have larger $Mg/Fe$. More recently Trager et
al (2000b) showed that there is a clear correlation between the ratio E/Fe
\footnote{''E'' refers to the total mass fraction of the elements
whose abundance is enhanced in giant elliptical galaxies with respect to Fe;
this group is similar but not exactly the same as the $\alpha$-elements group
(for more details see Trager et al 2000a).}
within $r=r_e/8$ and the velocity dispersion
$\sigma$ in a significant sample of local ellipticals. Their analysis,
extended also to other properties, favours a scenario in which local
ellipticals are composed by an old population encompassing $\sim 80 \%$ of
the stars and by a $20 \%$ of 'young' stars. In order to explain the
E/Fe-$\sigma$ correlation and the Z-plane (linking
$\left[Z/H\right]$, $\sigma$ and age), two possibilities have been
suggested: either  the IMF flattens with increasing $\sigma$, or $\alpha$
elements are more effectively ejected in smaller galaxies due to stronger
early winds. The two possibilities are not mutually excluding.

By converse, several authors suggested that in order to reproduce the
[E/Fe]-$\sigma$ correlation with reasonable parameters for the IMF and for
the SNIa rate and yields, the time scale for the burst must be short
($\sim$0.5-1 Gyr) for larger masses (see e.g. Matteucci 1994; Thomas,
Greggio and Bender 1999).

In the following we will test a scenario in which the star formation
begins at $t_*$ and proceeds very rapidly in most massive ellipticals and,
at $t_{QSO}$, the QSO stops the star formation by heating and ionizinig
the ISM and/or by triggering winds. Afterwards, only minor episodes  of
star formation can occur, in the sense that they involve $\lesssim$20$\%$
of the mass. In order to obtain the observed dependence of E/Fe and Z/H on
mass the star formation time scales are shorter (or conversely SFRs are
larger) for more massive galaxies.

We adopt for the duration in Gyr of the star formation burst
$T_{burst}= t_{QSO}-t_{\star}$ the following parametrization:

\be 
T_{burst}(M_{sph})=\left\{ \begin{array}{ll}
T_{b}^{*}, & \mathrm{if}\, M_{sph}\geq M^{*}_{sph}\\
T_{b}^{*}+ \log \frac{M^{*}_{sph}}{M_{sph}}, & \mathrm{if}\, 
M_{sph}\leq M^{*}_{sph}
\end{array}\right. 
\label{eq:delay1} \ee

\noindent  where $T_{b}^*=0.5$  Gyr is the burst timescale for galaxies with
$M_{sph}\gg M_{sph}^*$ ($M_{sph}^*\simeq 1.5 \times 10^{11} \ M_{\odot}$ is
the mass of an $L_{B}^*\simeq 3. \times 10^{10} \ L_{\odot}$ elliptical
galaxy). In the following we will refer to this parameterization as case A.
We also tested a case B, in which star formation lasts $T_{burst}=$0.5 Gyr
for all the spheroids, independently of the mass. It is worth noticing that
Monaco et al (2000), in order to account for the observed statistics of QSOs
and elliptical galaxies in the framework of hierachical structure formation,
introduced a a time delay decreasing with mass between the beginning of the
star formation and the QSO bright phase.

It is possible to explore whether the rates of star formation implied by equation
\ref{eq:delay1} are physically plausible in protogalaxies. In a dark halo with
circular velocity $V_c$ and mass $M_H \propto V_c^3$, the gas is fed to the  star
forming regions at a rate which is the minimum between the infall rate and the
cooling rate (White and Frenk 1991). The dynamical time  is
$\tau_{ff}(r)=\left(3\pi / 32 G \rho (r)\right)^{1/2}$. The cooling time
is defined as the ratio between the specific thermal content of the gas and the
cooling rate per unit volume
\be \tau_c(r)=\frac{3} {2}
\frac {\rho_g(r)} {\mu m_p} \frac {kT} {n_e(r)^2\Lambda (T)}
\ee
where $\Lambda(T)$ is the cooling function, depending also on chemical abundance
(Sutherland and Dopita, 1993), and T is the gas temperature. 
The latter is given by $kT=1/2 \, \mu m_p V_c^{2}$, where $\mu$ is the
mean molecular weight of the gas and $V_c$ is the circular velocity of the DM
halo, whose mass is $M_H$. Using the relationship between the $V_c$, the
virialization redshift $z_{vir}$ and $M_H$ (White and Frenk 1991) we have
\be
T=35.9 \left( \frac{M_H} {3.3\times 10^5 M_\odot} \right)^{2/3} (1+z_{vir}) \; {\rm K}
\ee
We use $\mu$ appropriate for a fully ionized gas with 25\% helium in mass and
adopt in the following
$z_{vir}=3$ as a reference value. 
The total mass of baryons is
$M_{tot}=f_b M_H$. We adopted for the baryon fraction $f_b=0.13$.

In the following we adopt a density profile for the DM appropriate for a $\Lambda
$CDM cosmology (Navarro, Frenk and White 1997). Depending on  mass and density
profile, we define a radius $r_{cool}$  inside which the gas can cool and collapse
on timescales $T_{burst}$. We denote by $f_{cool}$  the fraction of baryons within
this radius.

In order to explore the SFR, we use the following equation for the evolution
of the mass of cool gas $M_{cg}(t)$
\be \dot M_{\rm cg}(t) =- (1-R) \frac{M_{cg}(t)}{ \tau_{\star}}
- \dot M_{\rm w}
+C \exp (-\frac{t}{\tau_{inf}}). 
\label{eq:sfr1} \ee
\noindent 
The first term on the r.h.s. describes the cool gas converted into stars on
a timescale $\tau_{\star}$ and the restitution of gas
from stars.  We
assume that $\tau_{\star}=\max\left(\langle \tau_c\rangle,
\langle \tau_{ff}\rangle \right)$, where the two times are computed using
the average density inside $r_{cool}$, the limiting radius of the star
forming gas. 

The SFR=$M_{cg}/\tau_\star$ is expected to be affected by SN feedback.
However the fraction of energy transferred to the gas by the SNae is
rather uncertain. The second term in equation \ref{eq:sfr1} is a plausible
description of this feedback (Kauffmann, Guiderdoni and White 1994),
and represents
the rate of production of 'warm' gas that can not form stars
immediately
\be \dot M_{\rm w}(t) =\frac{4}{5} \frac {\nu f_{h}
E_{SN} {\rm SFR}} {V^2_c} , \label{eq:sfr2} \ee
where $\nu$ is number of SN produced per unit mass of formed stars. We adopt
here and in the following a Salpeter initial mass function (IMF),
$\Phi\propto M^{-x}$ with $x=1.35$, between $M_{l}=0.15 M_{\odot}$ and
$M_{up}=120 M_{\odot}$ and a supernova  mass threshold $M_{SN}\geq 8 \
M_{\odot}$. With these assumptions $\nu \simeq 8.7 \times 10^{-3} \,
M_{\odot}^{-1}$. We set the fraction $f_h$ of the SN energy output ($E_{SN}
\simeq 10^{51}$ ergs) that is transferred to the gas to $0.15$; of course the
results for low masses are rather sensitive to the adopted value of $f_h$.

The last term in equation \ref{eq:sfr1} describes the rate at which cold gas
is supplied, namely the 'infall' rate. As for the timescale in the
exponential law, we use $\tau_{inf}(r_{cool})=
\max\left(\tau_{c}(r_{cool}),\tau_{ff}(r_{cool})\right)$. The constant
$C=M_{tot}/  \tau_{inf}$ is the initial rate.

\begin{figure}
\psfig{file=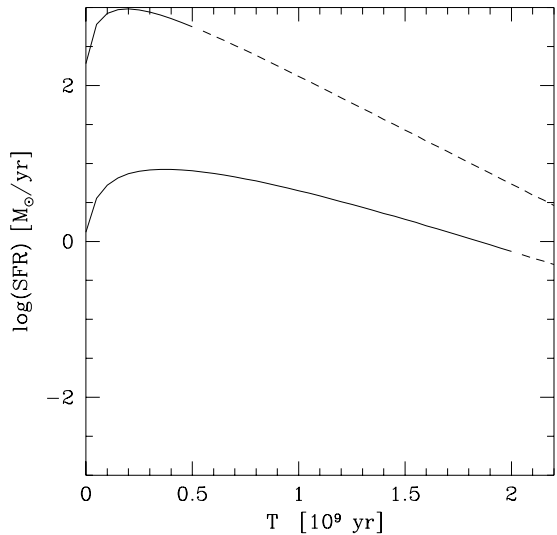,width=8.5truecm}
\caption{
Star Formation Rate for spheroidal galaxies with halo mass $M_{H}=
1.5 \times 10^{13} \ M_{\odot}$ and final mass in stars
$M_{sph}=4.3 \times 10^{11} \ M_{\odot}$ (upper curve) and
with $M_{H}=
1 \times 10^{11} \ M_{\odot}$ and 
$M_{sph}=5 \times 10^{9} \ M_{\odot}$ (lower curve).
The change from solid to dashed line marks the end of the star burst
at $T_{burst}$ given by equation \ref{eq:delay1}, because of the QSO shine.
In the paper we assume that the SFR vanishes for $T > T_{burst}$.
}
\label{fig:sfrgig}
\end{figure}

In Fig.~\ref{fig:sfrgig} the SFR, estimated using equation \ref{eq:sfr1},
are presented for spheroids with halo mass $M_{H}= 1.5 \times 10^{13} \
M_{\odot}$ and $M_{H}= 1 \times 10^{11} \ M_{\odot}$. The final masses in
stars are $M_{sph}=4.3 \times 10^{11} \ M_{\odot}$ and $M_{sph}=5 \times
10^{9} \ M_{\odot}$ respectively. It is apparent that the SFR can be as
high as 1000 $M_{\odot}$ in most massive spheroids. In these cases the
fraction of the baryons involved in the star forming process is
$f_{cool}\simeq 0.2$, since the gas in the outer regions has longer
cooling time. At small masses $f_{cool}=1$, but the effect of SN feedback
keeps the SFR low and prolongues the star formation, despite the shorter
cooling times. 

We exploited equation 11 to compute the SFR for objects with final mass in
stars ranging from $10^9$ to $5\times 10^{11}$ $M_{\odot}$. The SFR, averaged
over the time $T_{burst}$, as function of mass turns out to be

\be
SFR\simeq 100 \left(M_{sph}/
1 \times 10^{11} M_\odot\right)^{1.3} \ M_{\odot}\ yr^{-1}.
\ee

Only about 20-40\% of the baryons are locked in stars, the rest being in
form of warm gas. For large hosts only a small fraction of the gas has a
cooling time short enough to collapse before the QSO reionizes, reheats
and possibly expels the gas, stopping both star formation and accretion on
the black hole. For small mass hosts, star formation cannot proceed much
before the QSO shining because of the efficiency of SN feedback. As a
result the stars are again a small fraction about 3-5\% of the DM mass.

We checked that the average stellar metallicity increases with the mass.
Large SFR in massive hosts rapidly increases the metallicity of the gas
and forming stars  in such a way that it is about
solar before the QSO advent. Conversely in small hosts, the SNe feedback
keeps the average SFR low before the QSO blocks the star formation. Thus,
although the QSO advent is delayed, nevertheless a lower metal fraction is
produced. In our scenario SNIa explode in large hosts after the star
formation and QSO output prevent the gas from forming stars. For
less massive objects still important episodes of star formation can occur
even after the SNIa begin to explode. This produces a positive E/Fe-$\sigma$
correlation, as observed. The slope of the correlation significantly depends
on SNIa progenitors and the ensuing timescale for Fe enrichement (Thomas,
Greggio and Bender 1999; Kobayashi, Tsujimoto and Nomoto 1999; Nomoto et al
1999). The details of the chemical evolution will be discussed elsewhere
(Romano et al.\ in preparation).

The rate at which spheroids of present day mass $M_{sph}$ begin to form stars at
$t_{\star}$  is directly related to the rate  at which they appear to host a QSO
(equation 6 and Fig.~\ref{fig:bulges}) at time $t_{QSO}=t_{\star}+T_{burst}(M)$.

\section {The spectral evolution of spheroidal galaxies and their far--IR emission}
\label{sec:sed}

In the previous section we computed the cosmological rate at which spheroids
begin their star formation activity. We also estimated how the SFR depend on
$M_{sph}$. To compare with observations we need as well predictions about
their SED evolution. 

The galaxy surveys at submillimeter wavelengths with SCUBA and analogies
with local starburst galaxies suggest that the starburst phase occurred in a
dusty environment. The spectrophotometric evolution in the UV and optical
bands of bursts of star formation, under the assumption of no or negligible
dust absorption, has been investigated by many authors (e.g.\ Bruzual 1983;
Arimoto \& Yoshii 1986; Bruzual \& Charlot 1993; Bressan, Chiosi \& Fagotto
1994). While the no dust approximation well describes the physical situation
in elliptical galaxies at later times, dust is expected to play an important
role during the starburst epoch. The main effect of dust is to transfer a
major fraction of the emitted power from the UV and optical bands to the
mid- and far-IR.

In order to compute the SEDs of the spheroids with vigorous star formation
and dust, we use the model GRASIL developed by Silva et al.\ (1998), which
is state-of-the-art as far as dust reprocessing is concerned. GRASIL has
been tested against UV to radio SEDs of local spirals and starburst galaxies
(Silva et al.\ 1998, Silva 1999). It has also been used to reproduce the IR
properties of local galaxies in the framework of semi-analytical models
(Granato et al 2000). GRASIL includes: ({\it i}) chemical evolution;
({\it ii}) dust formation, assumed to follow the chemistry of the gas; ({\it
iii}) integrated spectra of simple stellar populations (SSP) with the
appropriate chemical composition; and ({\it iv}) appropriate distribution of
stars, molecular clouds (in which stars form and subsequently escape) and
diffuse dust. In particular, dust is spread over the whole galaxy and its
temperature distribution is determined by the local radiation field. It is
noteworthy that the assumption of a single dust temperature turns out to be
a significant oversimplification.

The predictions of sub-mm fluxes are plagued by the uncertainty of the
far-IR wavelength dependence of the dust emissivity which, {\it convolved
with the temperature distribution of dust grains}, determines the sub-mm
steep slope of the SED.  Longward of $\sim 40 \mu$m the dust opacity can be
represented by a power law $k_{\nu}\propto \nu ^{\gamma}$.  Values ranging
between 1 and 2 are commonly adopted for the spectral index $\gamma$. The
classical computations by Draine and Lee (1984) predict $\gamma=2$, but
there are several indications of a shallower slope at least in some cases.
The issue is further confused by the fact that in most papers the observed
sub-mm continuum is fitted with single temperature gray bodies.  However,
Silva et al.\ (1998) found that the IR spectrum of the archetypal
ultraluminous infrared galaxy (ULIRG) Arp 220 is well represented by
$k_{\nu}\propto \nu ^{1.5}$, whilst $k_{\nu}\propto \nu ^{2}$ is more
suitable for M82, the prototype starburst galaxy, and other galaxies. Arp
220 may however be more representative of a spheroid. Since a single opacity
law may be too strong an assumption we treat $\gamma$ as a free parameter
allowed to vary between 1 and 2.

\begin{figure}
\psfig{file=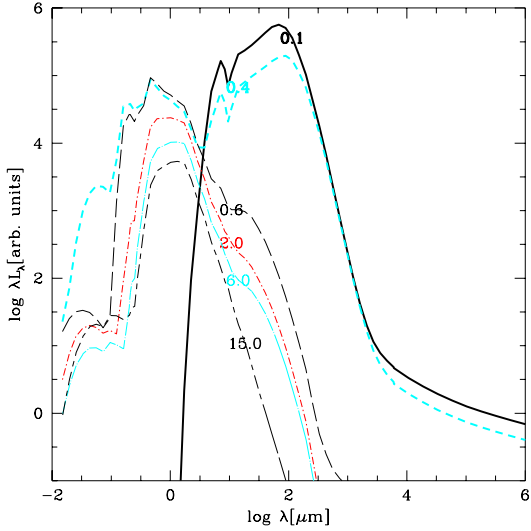,width=7.5truecm}
\caption{
Model SEDs of elliptical galaxies before (thicker lines) and after (thinner
lines) the QSO activity for $T_{burst}=0.5$ Gyr.
Numbers along the curves are the age in Gyr of the corresponding models.
}
\label{fig:sed05}
\end{figure}

\begin{figure}
\psfig{file=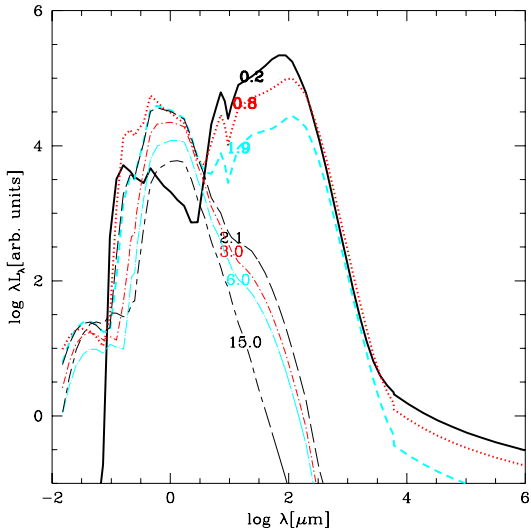,width=7.5truecm} 
\caption{Same as the previous figure but for $T_{burst}=2$ Gyr }
\label{fig:sed2}
\end{figure}

In Figs.~\ref{fig:sed05} and ~\ref{fig:sed2} model SEDs for elliptical
galaxies are shown as a function of time adopting  $\gamma= 1.5$. In the
former case, gas and dust are removed and star formation is stopped at
$T_{burst}=0.5$ Gyr (correspinding to the SFR in the upper curve of
Fig.~\ref{fig:sfrgig}), while in Fig.~\ref{fig:sed2} $T_{burst}=2$ Gyr
(lower curve of Fig.~\ref{fig:sfrgig}). It is worth noticing that GRASIL
predicts that the fraction of bolometric luminosity emitted in the far--IR
decreases with time during the burst phase. In the case depicted in
Fig.~\ref{fig:sed2} this fraction declines from about 90\% to about 40\%.

At the end of the burst,  gas and dust are removed and the stars already
formed emerge and shine in the optical bands, while the far-IR luminosity
drops dramatically. Of course both cases exhibit the same behavior at
$T>>T_{burst}$. However, the model with $T_{burst}=0.5$ Gyr passes through a
relatively blue phase (in the rest frame) during the first Gyr after the
onset of the galactic wind, while in the case  $T_{burst}=2$ Gyr, the galaxy
emerges with redder colours.

The UV and optical luminosities of the spheroids during the starburst dusty
phase critically depend on the relative distribution of stars and dust,
which sets the small fraction of UV luminosity escaping from molecular
clouds (where stars are born).  Therefore any model has large uncertainties,
and only observations can enlighten the physical processes involved. For
instance, local starburst galaxies, which are both strong IR emitters and
relatively blue objects, are well reproduced by GRASIL, allowing stars to
emerge from rather thick clouds ($\tau (1 \mu m)\sim 30$, as typical in the
Galaxy) on timescales of the order of $10^7$ yr (Silva et al.\ 1998).
Conversely Silva (1999) has shown that a very red optical spectrum,
resembling that of the Extremely Red Object HR 10 (Hu \& Ridgway 1994;
Cimatti et al.\ 1998; Dey et al.\ 1999), can be obtained assuming longer
timescales for the escape of stars from molecular clouds, endowed with much
lower optical thickness ($\tau (1 \mu m)\sim 2$). On the other hand the SED
predictions are more robust in the far--IR, the most important uncertainty
beeing related to the dust opacity spectral index $\gamma$ discussed above.

The radio emission is computed starting from the type II SN rate (Condon \&
Yin 1990), under the assumption that all stars with mass $M_{SN}\geq 5\
M_{\odot}$ end in a SN; the radio luminosity has to be decreased by a factor
$\sim 2$ if $M_{SN} \geq 8\ M_{\odot}$ is assumed (see Silva 1999 for
details).

\section {The submillimeter counts and the duration of the
burst in spheroids}
\label{sec:tes}

\begin{figure}
\psfig{file=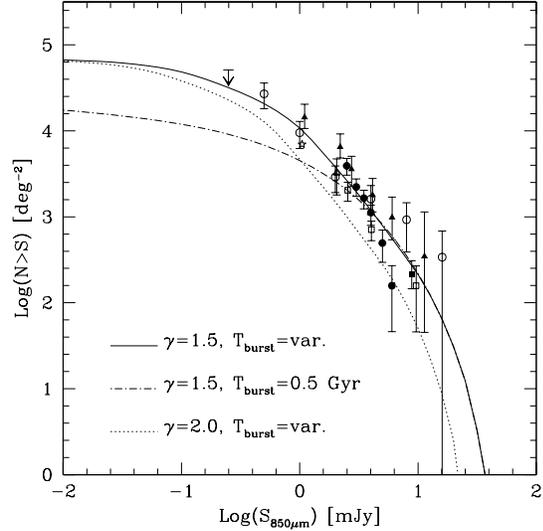,width=7.5truecm}

\caption{Predicted contributions of the spheroids to 850 $\mu$m
counts. Solid line is for case A ($T_{burst}$ given by equation
\ref{eq:delay1}) and $k_\nu\propto \nu^{1.5}$) while dot-dashed
line is for case B ($T_{burst}=0.5$ Gyr and $k_\nu\propto
\nu^{1.5}$). Data are from Blain et al.\ (1999b) (open circles),
Hughes et al. (1998) (star and open triangles), Barger, Cowie and
Sanders (1999) (open squares) Eales et al. (2000) (filled
circles), Chapman et al. (2000) (filled triangle) Borys et al.
(2000) (filled squares).}
\label{fig:850}
\end{figure}

In our scheme we identify the observed galaxies with $S_{850}\gtrsim 1$ mJy
with the dusty starbursts, occurring in spheroidal galaxies before the
shining of the QSO. From Figs.~\ref{fig:sed05} and \ref{fig:sed2} it is
apparent that the duration $T_{burst}$ of the dusty starburst can be tested
by far-IR counts and related statistics.

Coupling the SED predicted by GRASIL with the formation rate of the
spheroids $\dot n_{\star}(M_{sph},t_{\star})$, we compute the expected
contribution of spheroids to the submillimeter galaxy counts. The parameters
mainly affecting the fit to observed counts are the spectral index of dust
emissivity $\gamma$ and $T_{burst}$.

With the assumption $T_{burst}=0.5$ for all masses, the spheroids would
reproduce 850 $\mu$m counts only at brightest fluxes, allowing for a
substantial contribution from an additional population of possibly lower
redshift objects. By converse with a burst duration depending on the mass as
in case A (equation \ref{eq:delay1}) we obtain an excellent fit to the
observed counts with $\gamma=1.5$, as shown in Fig.~\ref{fig:850}. Therefore
in the following we concentrate on this solution, unless otherwise
explicitly stated.

It is worth noticing that also a longer duration of the burst $T_{burst}=2$
Gyr, with a corresponding decrease in SFRs, could give an acceptable fit to
the counts, provided that $\gamma=2$ is adopted. Howevever this solution is
not satisfactory, since it would decrease the the $Mg/Fe$ ratio below the
average observed value for local elliptical galaxies $Mg/Fe\simeq 0.2$
(Kobayashi and Arimoto 1999; Trager et al. 2000 a,b).

\begin{figure}
\psfig{file=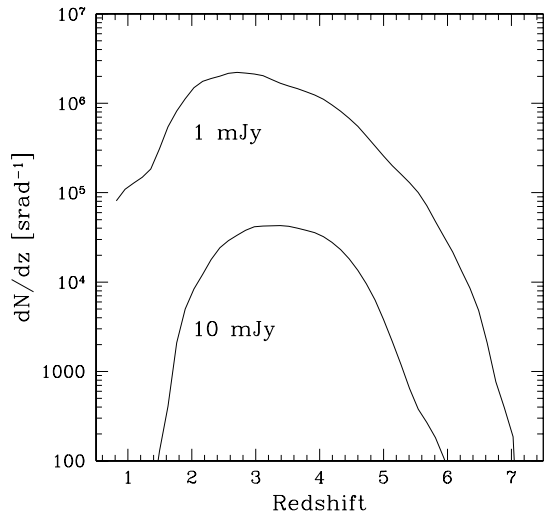,width=7.5truecm} 
\caption{
Predicted redshift distribution of spheroids  brighter
than 1 mJy and 10 mJy at 850 $\mu$m for case A.}
\label{fig:dis}
\end{figure}

Since in our scheme the infrared emission  in the spheroids occurs before
the QSO shine, the SCUBA sources must be located at high redshift. For
fluxes $S_{850}\gtrsim 1$ mJy, the predicted redshift distribution 
is relatively broad, with almost all the sources lying in the redshift
interval $2\lesssim z \lesssim 5$ (Fig.~\ref{fig:dis}). However we predict a
non negligible number of sources at $z\gtrsim 5$. Even at a flux limit
$S_{850}\gtrsim 10$ mJy almost all the sources are predicted to lie at
$z\gtrsim 2$. This is in agreement with the results obtained so far by the
spectroscopy of the optically identified counterparts of SCUBA sources
(Blain et al.\ 1999c; Smail et al.\ 2000). A testable prediction of the
proposed scheme is that there are more  extragalactic sources with
$S_{850}\gtrsim 1$ mJy at $z\geq 5$ than at $z\leq 1$, where only $\lesssim
1\%$ are located. 

The model source counts at $S_{850}\gtrsim 1$ mJy are dominated by the
massive spheroids (E/S0 galaxies), which have presently large luminosities
$M_B\leq M_{B\star} \simeq -20.5$.

How do the properties of the starbursting spheroids compare with those of
the individual submillimeter galaxies? 
The estimates of SFR for SCUBA galaxies range from several hundreds
to a few thousands $M_{\odot}
\mbox{yr}^{-1}$ (Ivinson et al 2000; Frayer et al 2000)
Two SCUBA sources have been
optically identified with Extremely Red Objects.  The available optical
identifications point toward very faint ($I\ge 26-27$) counterparts (Smail
et al.\ 1999). In our model, the most massive spheroids have SFR ranging
from several hundred to thousand $M_{\odot} \mbox{yr}^{-1}$ and they are
faint  at UV and optical wavelengths during the dusty phase. However, it
is worth mentioning that the expected optical magnitudes and colours of
these high z objects are very sensitive to the tiny fraction of UV
emission surviving after dust extinction and can in practice span a large
range of values, even at a fixed redshift. For instance,  we can reproduce
the 850 $\mu$m source counts with objects  exhibiting, during the dusty
phase, colours similar to those of EROs, as well as with relatively bluer
ones. This can be achieved simply by tuning the small fraction of young
stars which have already moved out from the parent clouds and/or on the
typical optical depth of clouds (cfr. sect. 3 and 5).

The extraordinary bursts of star formation before $t_{QSO}$ result also in
large radio luminosities.  Our scenario predicts that the
850 $\mu$m sources with $1\lesssim S_{850} \lesssim 10$ mJy have radio
fluxes at 5 GHz in the range  $5 \lesssim S_{5}\lesssim 200$ $\mu$Jy,
in agreement with the first results on the  radio 
emission of  SCUBA galaxies obtained by Smail et al.\ (2000)
and with the findings of Barger, Cowie and Richards (2000).

At 450 $\mu$m Blain et al. (1999c) report $N(>S=10\ {\rm mJy})=2.1\pm 1.2
\times 10^3 \ {\rm deg}^{-2}$. We predict that at this flux level the dusty
spheroids are $0.7 \times 10^3 \ {\rm deg}^{-2}$, a significant fraction  of
the counts. By contrast, at 175 $\mu$m the expected contribution is
negligible at the bright limit $S_{175}\sim 100$ mJy (Puget et al.\ 1999;
Kawara et al.\ 1998). This is due to the fact that the K-correction is not as
favourable as it is at longer wavelengths, dimming sources at $z\gtrsim 3$,
where most of the dusty spheroids are  located (cfr.\ Fig.~\ref{fig:dis}).
Indeed Scott et al.\ (2000) have observed with SCUBA at 450 and 850 $\mu$m a
subsample of the objects selected at 175 $\mu$m. The resulting submillimeter
spectral energy  distributions suggests that these objects are at low or
moderate redshift $0\leq z\leq 1.5$.

Going to shorter wavelengths, the spheroids  emerge as an important
component below 10 mJy at 60 $\mu$m;  at 90 $\mu$m, they  start to be a
non negligible fraction at fluxes $\lesssim 20$ mJy. The latter limit is
within the reach  of the ELAIS survey, covering about 20 square degrees at
90 $\mu$m (Efstathiou et al.\ 2000).

Several authors presented models of galaxy evolution taylored to fit
available data and in particular the submillimeter counts. For instance
Guiderdoni et al (1998) reproduced  the counts by imposing a large evolution
in the fraction of mass density involved in bursts of star formation and in
the fraction of Ultra Luminous Infrared Galaxies (ULIGs), both fractions
passing from a few percent in the local Universe to substantial values at
high redshift. The conclusion that rapid star formation activity is required
at $z \gtrsim 3$ has been claimed by Blain et al.\ (1999a). They showed
that, in order to reproduce the observed sub-mm counts, the activity in
galaxies had to be a factor of 200 larger at $z\sim 3$ than in the local
Universe. In their scheme this activity refers to both starbursts and to
AGNs, and is triggered by merger events at high redshift in a hierarchical
model of galaxy formation. A view even closer to our own has been proposed
by Tan, Silk \& Balland (1999). They exploited a model in which the final
morphology of the galaxies depends on the number of collisions and tidal
interactions they suffer (Balland, Silk \& Schaeffer 1998), deriving for
spheroids large star formation rate density at high redshift. In particular,
they broadly reproduce the 850 $\mu$m counts with objects that suffered
important collisions and that they identify as precursors of spheroidal
objects.

Our model is characterized by a very steep increase of the counts below
$\simeq 60$ mJy. Above this flux limit the sources are mainly low redshift
spirals and starburst galaxies, while just below it a large amount of high
redshift star forming spheroids appear.

In conclusion the most massive spheroids bursting at $z\gtrsim 2$, before
the onset of the QSO activity, can be identified with the observed SCUBA
sources down to 1 mJy. The SCUBA galaxies are strongly clustered, since they
are shining in the most massive virialized haloes at high redshift. The
large rate at which the spheroids form at high redshift, is dictated by the
QSO luminosity function and by the short QSO lifetime $\Delta t_{Q}$. The
other relevant parameter is the duration of the star formation,
observationally bounded also by the chemical evolution of QSO hosts and in
general of elliptical galaxies. The implied star formation rates are
physically plausible in protogalaxies (cfr Sect. 2.2).

\section {Starbursting and post-starburst spheroids and Lyman Break galaxies}
\label{sec:lbg}

In the context of the star formation in spheroids at high redshift, a
relevant issue is the relation of high redshift galaxies selected in optical
and near IR bands with the submillimeter selected galaxies. Steidel et al.\
(1999), using a galaxy sample complete to $I_{AB}\lesssim 25$ and
appropriate color criteria, were able to identify a large sample of galaxies
with the Lyman Break falling in the optical or near UV bands. These objects
at $z\ge 3$ have been claimed to exhibit dust absorption with reddening in
the range $0\leq E(B-V) \leq 0.4$, with an average value of $\sim 0.15$
(Steidel et al.\ 1999). After correction for the implied dust extinction,
the UV luminosity density of the LBGs at $z\sim 3-4$ would be almost equal
to that inferred from galaxies lying at $z\sim 1$ (Connolly et al.\ 1997).
The corresponding extinction corrected (by a factor of about 5) star
formation rate of a typical $M_{\star}$ galaxy would be of about $60 \
M_{\odot} \mbox{yr}^{-1}$. This implies a SFR per unit volume almost
constant $\dot\rho_{SFR}\simeq 0.15 \ M_{\odot}\ \rm yr^{-1} \ \rm Mpc^{-3}$
from $z\sim 1$ to $z\sim 4$.

On the other hand SCUBA observations of a selected subsample of LBGs
resulted in only one possible detection, suggesting that the SFR of the
population as a whole is possibly low (Chapman et al.\ 1999). However this
result is expected in our scheme, because of selection effects (see also
Adelberger and Steidel 2000; Peacock et al. 2000). The UV luminosity during
a star burst critically depends on the fraction of young stars already
escaped from their parent molecular clouds and on the associate time scale
(cfr. Section 3 and Silva et al.\ 1998). We already remarked that it is more
prone to model uncertaities than the IR and sub-mm emission. In any case,
the typical behavior of our models is shown in Fig.~\ref{fig:bulges} and
Fig.~\ref{fig:sfrgig}. The ratio $L_{FIR}/L_{UV}$ varies from 5-10 at about
the end of the burst to very large values just after the begin. After the
QSO phase the ratio falls down very rapidly, in the short period when the
spheroids are still blue unextincted objects.

\begin{figure}
\psfig{file=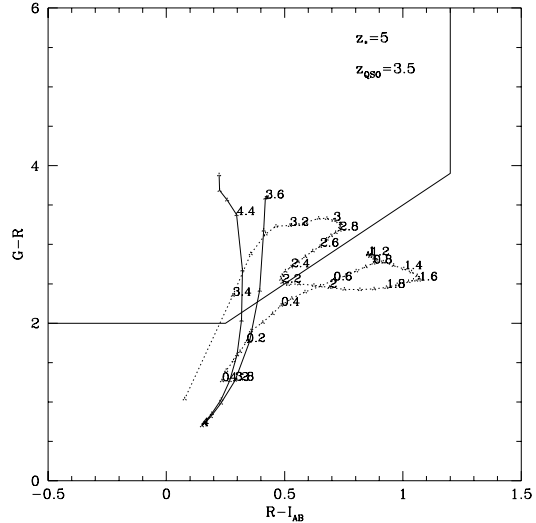,width=7.5truecm} 
\caption{Track of the model of Fig.\ 1, assumed to begin star formation at
$z_{\star}=5$, in the two colour plane $(G-\cal{R})$ vs.\ $(\cal{R}-I)$ used by
Steidel et al.\ (1999) to select  LBG candidates.  
The upper left region bounded by
solid lines is the selection region.
Number along the
track (solid line before the wind and dotted afterwards) are the corresponding redshifts.}
\label{fig:col}
\end{figure}

Keeping in mind the above mentioned caveats concerning the dusty
star-forming phase, our modelling predicts that the spheroids may spend most
of their $z\geq 3$ life inside the region of the colours plane used to
select the LBG candidates (Steidel et al 1999). For instance, a massive
galaxy $M_{sph}\simeq 3\times 10^{11} \ M_{\odot}$ with star formation
started at $z_{\star}\simeq 5$ and with the QSO activity appearing after
$T_{burst}\simeq 0.5 \ Gyr$, would have $I_{AB}\lesssim 25$ since $z\lesssim
4$, and would exhibit colours $(G-\cal{R})$ and $(\cal{R}-I)$ quite similar
to those of the LBGs, as evidenced by Fig.~\ref{fig:col}.

In our scenario the hosts of bright QSOs ($L_{bol}\geq 10^{13}  \ L_{\odot}$
may appear as LBGs both during their starburst phase and  soon after the QSO
shining; these hosts now have luminosities $M_B\lesssim -20.5$ and velocity
dispersion $\sigma \simeq 200$ km/s. As stated in section 4, these hosts are
also seen as SCUBA objects during their starburst, since they have average a
SFR in the range 100-1000 $M_{\odot}$ yr$^{-1}$. The metal abundance  of
their stars, estimated with a Salpeter IMF, is about solar. These massive
spheroids are also expected to exhibit a large rate of SNIa explosion, after
the QSO shining, when they are almost gas and dust free.

Less massive spheroids, with present luminosities $-17\leq M_B\leq -20.5$,
at redshift $z\sim 3$ have average SFR from a few to one hundred $M_{\odot}\
yr^{-1}$ and sub--solar metal abundance in stars, decreasing with decreasing
SFR. The majority of these spheroids have not yet experienced the QSO
shining at $z\geq 3$ and they are in the pre-QSO phase. In our modelling of
the SED (cfr. sect 3) dust follows the gas chemistry and, as a consequence,
the extinction is expected to decrease with decreasing SFR. Thus the less
massive spheroids can appear as LBGs and may constitute a major fraction of
them.  In conclusion the spheroids independently of their mass very likely
showed up at $z\geq 3$ as LBGs. However they have SFR, metal abundance, dust
extinction and clustering scale decreasing with their mass.

It is also worth noticing that in our schematic  model star formation has
completely stopped after $t_{QSO}$. However, ionization, heating and winds
due to the QSO can simply decrease the SFR  from several hundred $M_{\odot}$
yr$^{-1}$ to several $M_{\odot}$ yr$^{-1}$, compatible with the observed UV
emission from a typical LBG. Since this more prolonged star formation would
involve only a minor fraction of the galaxy mass, say 5-10 per cent of the
mass in 1-2 Gyr, the present day spectra of the galaxy would be unaffected.
A residual low level of star formation activity has been inferred from broad
band spectra of the HDF elliptical galaxies at $z\sim 1$ (Franceschini et
al.\ 1998).

\section {The QSO phase}
\label{sec:qso}

The QSO can heat, ionize and also expel the ISM, thus strongly
decreasing or even stopping the copious star formation in the host galaxy
at $t_{QSO}$ (Silk and Rees 1998; Fabian 1999). However still significant
star formation can be  present during  the QSO lifetime $\Delta t_{Q}\sim 4
\times 10^7$ yr. Several objects with these characteristics have already
been found (e.g.\ Omont et al.\ 1996; Ivison et al.\ 1998; Benford et al.\
1999; Yun et al.\ 1999). If we assume that the star formation is not
quenched during the nuclear activity, then the fraction of QSOs in an
unbiased submillimeter survey depends on the ratio between the QSO lifetime
and the burst duration  $T_{burst}$. Since the latter timescale in our
scheme is a function of the mass, this precise fraction depends on the flux
limit. In particular for bright submillimeter sources $S_{850}\gtrsim 2$ mJy
we expect $T_{burst}\simeq 0.5-1$ Gyr and then a fraction of QSOs from a few
to 10$\%$ is estimated. Recent observations of SCUBA galaxies with Chandra
suggest that they are mainly powered by starbursts, unless Compton-thick
tori with little circumnuclear X--ray scattering are in place (Fabian et al
2000; Hornschemaier et al 2000; Severgnini et al 2000). A rather different
result has been obtained by Bautz et al (2000), who were able to detect the
X-ray counterparts of two submillimeter sources in A 370. These authors
conclude that $20^{+30}_{-16} \%$ of the submillimeter galaxies exhibit
X-ray emission from AGN. Spectroscopy revealed no QSO among the 15 850
$\mu$m selected sources, though several of them exhibit possibile signs of
low nuclear activity (Ivison et al.\ 2000).

On the other hand hard X-ray selected sources down to $S_{2-10 keV} \sim 2-4
\times 10^{-15}$ ergs s$^{-1}$ cm$^{-2}$ are basically not detected even in
deep submillimeter exposures $S_{850}\gtrsim 2 $ mJy (Barger et al 2000;
Fabian et al 2000; Severgnini et al 2000), with a few exceptions. In
particular in the sample of Barger et al (2000) most of the X-ray
contribution to the resolved X--ray background comes from AGN with
$L_{bol}\sim 10^{45}-10^{46}$ erg s$^{-1}$, associated to BH masses
$M_{BH}\sim 10^7-10^8 \ M_{\odot}$ and baryon masses in spheroids $M_{sph}\
\sim 5\times 10^9-5\times 10^{10} \ M_{\odot}$, in keeping with the
predictions of Salucci et al (1999).  At 850 $\mu $m, contributions from
dusty tori around the active nucleus and from star bursts in host galaxies
are expected.  However the dust emission from the torus peaks at
$\lambda\simeq 30 $ $\mu$m, though in case of very extended structures a
wide spread of temperatures tends to produce a plateau up to 80 $\mu$m
(Granato and Danese 1994; Efstathiou \& Rowan-Robinson 1995).  Thus
relatively faint AGN at $z\sim 1-2$ can not be detected at $S_{850}\geq 1$
mJy. Also the starburst emission is elusive, because of the relatively low
star formation expected in these objects.  For instance in our model we
expect that their host galaxies have already formed most of their stars in
about 1-2 Gyr, with an average $SFR\sim 2.5-50 \ M_{\odot}\ yr^{-1}$. These
limits are below possible detection by SCUBA for galaxies at $z\geq 1$.
Then, as claimed by Granato, Danese \& Franceschini (1997), the obscured
AGNs, which produce a major fraction of the Hard X-ray Background Radiation
(HXRB), contribute only a small fraction of the FIRB at submillimeter
wavelengths (see also Gunn ands Shanks 1999).

Interestingly enough, the evolutionary links between the star bursting phase
and the nuclear activity we have envisaged at high redshift, can be applied to ultra
luminous infrared galaxies at low redshift (Sanders et al.\ 1988). In our scheme the 
low redshift ULIRGs
correspond to the dusty starburst phase. This is in keeping with recent ISO
observations that show that for the local ULIRGs, the starburst activity is
sufficient to power the bulk of the bolometric luminosity (Genzel et al.\
1998). However we also expect cases in which the QSO light reprocessed by
dust in the host galaxy is dominating over the far--IR emission due to the
starburst.

\section {Discussion}
\label{sec:dis}

\subsection {QSO hosts}
 
A major assumption in our model is that the relationship between BH mass and
host galaxy mass has been imprinted at early epochs (section 2.1). Recent
observations from ground and with HST show evidence that the radio loud QSO
and the radio galaxies at high redshift $z\geq 1$ are hosted in bright
spheroidal galaxies with a rather evolved stellar population (Dunlop 1999;
Zirm et al 1999; Lacy, Bunker and Ridgway 2000; McLure and Dunlop 2000).
These observations support our assumption. Also recent photometric and
spectroscopic follow-up of a hard X-ray selected sample showed that a large
fraction of the $z\gtrsim 1$ AGN, responsible for a major fraction of the
HXRB, are hosted in early type optically luminous galaxies (Barger et al
2000).

Observations of radio quiet QSO hosts show that their luminosity at fixed QSO
luminosity decrease with increasing $z$, (McLure at al 1999; Ridgway et al
1999; Rix et al 1999). The 5 QSOs with $z\geq 1.7$ studied by Rix et al
(1999) exhibit a median ratio of the nuclear to the host galaxy luminosity
$L_N/L_{host}\simeq 6$ in the V-band, while in the R-band for 9 low redshift
($z\leq 0.24$) QSOs   McLure et al (1999) found $L_N/L_{host}\simeq 1.5$.
This decrease by a factor 4 is explained by the observed decline of the
median Eddington ratio from $L/L_{Edd}\sim 1-2$  to $L/L_{Edd}\simeq 0.1-0.2$
for QSO at high redshift and low redshift respectively (Padovani 1989; Sun
and Malkan 1989; Wandel 1999). As a matter of fact the low redshift radio
quiet QSO of the sample of McLure et al (1999) exhibit a median
$L/L_{Edd}\simeq 0.2$, under the assumption that $M_{bh}/M_{sph} \simeq
3\times 10^{-3}$.

The observed decrease of the $L/L_{Edd}$ ratio is expected in our scenario.
When QSOs shine at high redshift $z\geq 1.5-2$ they have still large gas
reservoirs in the hosts and they can reach the self-regulated Eddington
accretion. At lower redshift, gas in hosts is scanty and only encounters with
other galaxies can supply material for re-activation. However this supply can
only occasionally be large enough to reach the self-regulated regime (see
also Cavaliere and Vittorini 2000). Note that the decrease of $L/L_{Edd}$ is
quantitatively reproduced by the semi-analytic model by Kauffmann and
Haehnelt (2000). In this case however it is driven not only by the lower gas
fraction left at lower redshift, but also by the adopted redshift dependence
of the accretion timescale $t_{acc} \propto (1+z)^{-1.5}$ in their best fit
model.

\subsection {Early baryon collapse in massive DM haloes}

The hierarchical scheme in the Dark Matter component can naturally explain
the onset of  QSO activity and of star formation in spheroids. In
particular the Press-Schechter formalism within the standard variants
predicts a large enough number of massive DM halos in order to host QSOs
and their parent galaxies (see e.g. Haehnaelt and Rees 1993; Haehnaelt,
Natarajan and Rees 1998; Cattaneo, Haehnaelt and Rees 1999; Monaco,
Salucci and Danese 2000). However a distinct character of hierachical DM
scheme is that smaller haloes virialize earlier. On the other hand the
high SFRs and IR luminosities implied by the far-IR counts exclude that
most of the stars have formed in smaller subclumps at very high redshifts,
say $z\sim 10$. It is plausible that at  high redshift $z\gtrsim 6$ the
SFR has been kept very low in the subclumps by 'internal' mechanisms
(e.g.\ SN feedback or  even a low activity of the nucleus), which became
progressively  ineffective when larger and larger masses virialized. Then,
as discussed in Sect. 2.2, a significant fraction of the baryons in
massive 'virialized' haloes can fragment very rapidly to form stars and
QSOs, while the star formation is further delayed in smaller haloes by the
effect of stellar feedback. The shining of the QSOs is the final and brief
time episode of the complex processes involved in assembling baryons in
stars and BH. Most of the relevant observational facts can be framed in a
scenario in which the most massive  objects complete their activity in a
very short time interval $T_{burst} \simeq 0.5$ Gyr and the less massive
in $\sim 2$ Gyr. In a sense the DM hierarchical sequence is corrected by
an {\it Anti-hierarchical Baryonic Collapse} (see also paper II),
regulated by the QSO and SN feedback.

Since star formation in massive ellipticals occurs early and in relatively
short times, the emerging picture is somewhat in between  the concept of a
'monolithic' collapse  of baryons inside the haloes and the  'classical'
hierarchical semi-analytic models,  discussed for instance by Kauffmann \&
Charlot (1998) and Governato et al.\ (1998). In particular, the connection
between ellipticals and quasars has been addressed recently by Kauffmann \&
Haehnelt (2000). In their 'classical' scenario of the merging process they
reproduce the QSO activity decline, but the  most massive spheroids form
late. As a result they predict a significant active evolution of the host
galaxy after the quasar shines. On the contrary in our scheme the host
galaxies evolve passively, except for subsequent minor mergers at lower
redshift which change the mass in stars by no more than 20\% (Trager et al.\
2000b; see also Van Dokkum et al.\ 1999).

It is very likely that the main aspects of the {\it Anti-hierarchical
Baryonic Collapse} can be reproduced by semi-analitycal models, by acting on
very basic parameters, such as merging and star formation timescales and SN
feedback efficiency.

\subsection {The star formation in spheroids}

\begin{figure}
\psfig{file=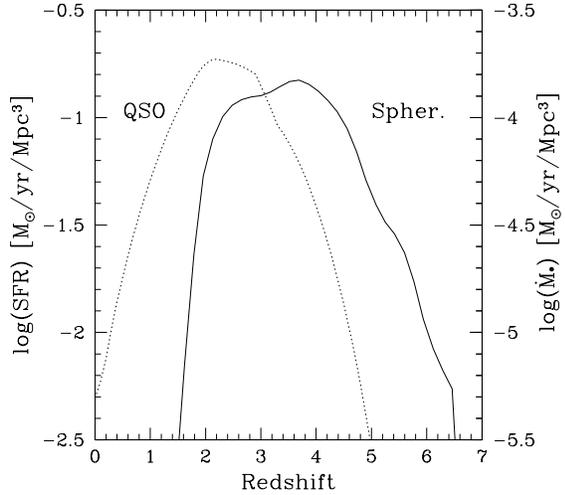,width=7.5truecm} 
\caption{
Star formation (solid line and data points, left scale) and mass accretion (dotted line,
right scale) rates per unit volume as functions of redshift.
The latter has been derived from the data on QSO luminosity function
evolution as a function redshift. The SFR refers to the Case A.}

\label{fig:sfr}
\end{figure}

The star formation rate per unit volume $\dot \rho_{\star}$ in
spheroids as a function of redshift derived from our modelling is
presented in Fig.~\ref{fig:sfr}. The mass accretion rate per unit
volume implied by the observed nuclear activity is also
presented. It is apparent that in our scheme the star formation
rate in QSO hosts is rapidly falling down at $z\lesssim 2$.
However the star formation rate in the Universe remains
significant because of the star formation in discs and dwarves.

The star formation rate inferred, without dust correction from the UV
emission of LBGs galaxies at $z\geq 3$ is a factor $\sim 5$ lower than our
prediction (Steidel et al 1999). However this is consistent with the
evidence of strong extinction of UV emission.

The density of mass cycled through stars in QSO hosts amounts to
$\rho_{c\star} \simeq 3.2 \times 10^8 \ M_{\odot} \ {\rm Mpc}^{-3}$, which
implies (using Salpeter IMF) a present density of long-lived stars in QSO
hosts $\rho_{\star}\simeq 2 \times 10^8  \ M_{\odot} \ {\rm Mpc}^{-3}$.
This mass density can be compared  to the mass density in stars in local
spheroids $2\times 10^8 \leq \rho_{\star}^{sph}\leq 5 \times 10^8 \
M_{\odot} \ {\rm Mpc}^{-3}$ (Fukugita et al 1999). The metal mass density
is roughly given by $\rho_Z \simeq y_Z  \rho_{c\star} \simeq 9 \times 10^6
\ M_{\odot} \ {\rm Mpc}^{-3}$, $y_Z$ denoting the fraction of metals
returned to the interstellar matter by a stellar generation (0.028 with
the adopted IMF). In our scenario only about 40\% of these  metals are
still locked into stars, while the rest are spread over the intergalactic
medium. The X--ray emitting gas in groups and clusters of galaxies amounts
to $\rho_{IGM}(z=0)\simeq 1\times 10^9 M_{\odot} \ {\rm Mpc}^{-3}$
(Fukugita et al.\ 1999). Available observations suggest that this gas has
$Z\simeq 1/3 Z_{\odot}$ (Renzini 1999), implying $\rho_{Z \
IGM}(z=0)\simeq 7 \times 10^6 \ M_{\odot} \ {\rm Mpc}^{-3}$, not far from
our prediction.

\section{Summary}
\label{sec:sum}

We have presented a unifying scheme for the formation of QSOs, elliptical
and S0 galaxies, and for the bulges of spirals. We have shown that the main
aspects of the evolution of spheroidal galaxies and QSOs can be well
understood if we assume at $t=t_{QSO}$, when the QSOs shine, the bulk of the
stars have already been formed in QSO hosts with star formation lasting a
time $T_{burst}=t_{QSO}-t_{\star}$ ranging from $\sim 0.5$ to $ \sim 2 $ Gyr
when going from  more to  less massive hosts. This suggest an {\it
Anti-hierarchical Baryonic Collapse}, i.e. that the baryons in large
spheroids collapse very rapidly to form stars and QSOs, while the collapse
of baryons in smaller spheroids is slowed down. In fact we have shown that
the central regions of the massive DM haloes can form stars on very short
timescales, while the SNe feedback is strongly slowing down the SFR in small
spheroids, extending the star forming phase. The ensuing formation rates
scale  as $SFR\simeq 100 \left(M/10^{11} M_\odot\right)^{1.3} \ M_{\odot} \
{\rm yr}^{-1}$. In our picture when the QSO reaches its maximum, the
combined actions of its power and SN feedback are able to practically stop
both the SFR and the nuclear accretion through ionization, heating and
winds. This influences  both the mass in stars $M_{sph}$ and the BH mass
$M_{BH}$ and may be the origin of the observed correlation between the two
masses. As a result only about 20-40\% of the baryons are presently in stars
both in massive and small DM galactic haloes. The remaining is in warm or
hot gas with $Z\simeq \frac {1} {3} Z_{\odot}$. The outlined scenario can
explain:

\begin{enumerate}
\renewcommand{\theenumi}{(\arabic{enumi})}

\item the main aspects of the chemical evolution of spheroids (stellar
metallicity, luminosity-metallicity relationship, the $\alpha$
enhancement) and the observed elemental abundances in QSOs;

\item the 850 source counts;

\item the main properties of Lyman Break Galaxies;

\item the detection of high redshift, red and old massive ellipticals.

\end{enumerate}

This general view implies the following evolution sequence for  QSO
hosts:

\begin{enumerate}

\item the most massive appear, at high redshifts $z\gtrsim 2-6$, as
ultraluminous far-IR galaxies (SCUBA galaxies). This phase  lasts for
$T_{burst} \sim 0.5-1 \ {\rm Gyr}$. On the average they are optically faint
$I_{AB}\gtrsim 26-27$ with colors difficult to predict, in the sense that,
depending on uncertain details of dust distribution, both objects with some
intrinsic UV emission as LBGs and very red objects (like EROs) can be
expected;

\item the hosts of intermediate and low mass have bursts more protracted
$T_{burst}\sim 1-2$ Gyr and at $z\geq 3$ can appear as LBGs;

\item the QSO phase then follows. The IR emission is  attributable both to
the starburst and the nuclear activity. This phase lasts for the QSO duty
cycle $\Delta t_Q \simeq t_{duty}\simeq 4 \times 10^7 \ yr$. It ends
because heating, and ionization of the ISM and  the possible onset
of galactic winds drastically reduce SFR and nuclear activity;

\item  a long epoch of passive
evolution follows, with spectra becoming quickly red.

\end{enumerate}

The scheme implies  that 850 $\mu$m selected galaxies with $S_{850}\geq 1$
mJy are strongly clustered. Also LBGs are expected to be significantly
clustered, though to a lower level.

The emerging picture is consistent with hierarchical models for structure
formation. However it evidences that, in order to understand key
observations of QSOs and elliptical galaxies, the hierarchical assembly of
DM haloes must be completed by the history of the baryons inside the haloes
themselves. The physical processes related to star formation and feedback
suggest that baryons can collapse in stars and QSO more rapidly in more
massive haloes. However why the QSOs shine in their hosts when star
formation has already progressed, is still an open problem.

\section*{Acknowledgements}
We thank Alessandro Bressan, Cesare Chiosi and Francesca Matteucci for
helpful discussions and David Borg for careful reading of the
manuscript. We also thank ASI and italian MURST for financial support.

\bsp

\label{lastpage}

\end{document}